\documentclass[aps,nofootinbib,floatfix,twocolumn,prl,pacs,class-pre]{revtex4}
\usepackage{amsfonts}
\usepackage{amssymb}
\usepackage{graphicx}
\usepackage{epsfig}
\newcommand{\be}{\begin{equation}}
\newcommand{\ee}{\end{equation}}
\newcommand{\bea}{\begin{eqnarray}}
\newcommand{\eea}{\end{eqnarray}}
\newcommand{\lapproxeq}{\lower .7ex\hbox{$\;\stackrel{\textstyle
<}{\sim}\;$}}
\newcommand{\gapproxeq}{\lower .7ex\hbox{$\;\stackrel{\textstyle
>}{\sim}\;$}}

\newcommand{\simlt}{\mathrel{\hbox to 0pt{\lower
      3.5pt\hbox{$\mathchar"218$}\hss} \raise
      1.5pt\hbox{$\mathchar"13C$}}}

\newcommand{\simgt}{\mathrel{\hbox to 0pt{\lower
      3.5pt\hbox{$\mathchar"218$}\hss} \raise
      1.5pt\hbox{$\mathchar"13E$}}}

\begin{document}
\title{Can TeVeS avoid Dark Matter on galactic scales?} \author{Nick
  E. Mavromatos}\email{Nikolaos.Mavromatos@kcl.ac.uk} \author{Mairi
  Sakellariadou} \email{Mairi.Sakellariadou@kcl.ac.uk}
\author{Muhammad Furqaan Yusaf} \email{Muhammad.Yusaf@kcl.ac.uk}
\affiliation{King's College London, Department of Physics, Strand WC2R
  2LS, London, U.K.}

\begin{abstract}
A fully relativistic analysis of gravitational lensing in TeVeS is
presented.  By estimating the lensing masses for a set of six lenses
from the CASTLES database, and then comparing them to the stellar
mass, the deficit between the two is obtained and analysed.
Considering a parametrised range for the TeVeS function $\mu(y)$,
which controls the strength of the modification to gravity, it is
found that on galactic scales TeVeS requires additional dark matter
with the commonly used $\mu(y)$. A soft dependence of the results on
the cosmological framework and the TeVeS free parameters is
discussed. For one particular form of $\mu(y)$, TeVeS is found to
require very little dark matter. This choice is however ruled out
by rotation curve data. The inability to simultaneously fit
lensing and rotation curves for a single form of $\mu(y)$ is a
challenge to a {\sl no dark matter} TeVeS proposal.

\vspace{.1cm}

\noindent
PACS numbers:
\end{abstract}

\maketitle


The standard ($\Lambda$CDM) cosmological paradigm is based upon Cold
Dark Matter (CDM), a cosmological constant $\Lambda$ and classical
general relativity/Friedman-Robertson-Lema\^{i}tre-Walker
cosmology. Despite its enormous success and consistency with a
plethora of astrophysical data, competing models have been proposed
for the primary reason of the still unknown nature of the dark energy
component and the current undetectability of dark matter. To explain
the observed flat rotation curves of galaxies without Dark Matter,
Milgrom~\cite{milgrom} proposed MOdified Newtonian Dynamics (MOND),
based upon the relation $f(|{\vec a}|/a_0){\vec a}=-{\vec
  \nabla}\Phi_{\rm N}$ between the acceleration ${\vec a}$ and the
Newtonian gravitational field $\Phi_{\rm N}$.  The constant
$a_0\approx 1.2 \times 10^{-10} {\rm m}/ {\rm s}^2$ is motivated by
the acceleration found in the outer regions of galaxies where the
rotation curve is flat.  When $f$, assumed to be a positive, smooth,
monotonic function, equals unity, usual Newtonian dynamics hold, while
when it approximately equals its argument, the deep MONDian regime
applies.

MOND has been successful in explaining the dynamics of disk galaxies,
however it is less successful for clusters of galaxies. It was
promoted~\cite{bek} to a classical relativistic field theory by
introducing a TEnsor, VEctor and Scalar field (TeVeS). TeVeS has been
criticised as lacking a fundamental theoretical motivation. Recently,
it has been argued~\cite{nickmairi} that such a theory can emerge
naturally within some string theory models.

In Ref.~\cite{fsy}, where the lensing mass in MOND was compared to the
stellar mass content of the lenses, it was found that comparable
amounts of dark matter were needed in MOND to that required in the
standard lensing scenario. This result is in contrast with attempts to
explain the lensing data on galactic-cluster scales by introducing a
2~eV neutrino. In fact, this component of dark matter has been
shown~\cite{sand07} to cluster on Mpc scales but not on galactic
scales where the previous analysis was conducted. It was concluded
that either lensing must operate in a qualitatively different way
within the covariant ``parent'' theory of MOND, such as the TeVeS
models, or dark matter should be considered within MOND even on
galactic scales. 

In this letter TeVeS is examined in a similar way,
namely by deriving the modified lensing equation and solving it
numerically, to investigate whether it also shows a dark component
from lensing. The (weak) dependence of the results on the
cosmological models and TeVeS free parameters is then
discussed. TeVeS~\cite{bek} is a bi-metric model in which matter and radiation
does not feel the Einstein metric, $g_{\alpha\beta}$, appearing in the
canonical kinetic term in the (effective) action, but a modified
``physical'' metric, ${\tilde g}_{\alpha\beta}$, related to the
Einstein metric by $\tilde{g}_{\alpha\beta} =
e^{-2\phi}g_{\alpha\beta}-U_\alpha U_\beta(e^{2\phi}-e^{-2\phi})$,
where $U_\mu , \phi$ denote the TeVeS vector and scalar
field, respectively. The TeVeS action is:
\begin{eqnarray}\label{teveslagr}
S=&&\int d^4x\ \left[\frac{1}{16\pi
    G}\left(R-2\Lambda\right)\right.\nonumber\\ &&-\frac{1}{2}\{\sigma^2
\left(g^{\mu\nu}
  - U^\mu U^\nu
  \right)\phi_{,\alpha}\phi_{,\beta}+\frac{1}{2}Gl^{-2}\sigma^4F(kG\sigma^2)\}
\nonumber\\ &&\left.-\frac{1}{32\pi
    G}\left\{K{\cal F}^{\alpha \beta} {\cal F}_{\alpha \beta}-2\lambda
  \left(U_\mu U^\mu + 1 \right)
  \right\}\right](-g)^{1/2}\nonumber\\ &&+{\cal
  L}(\tilde{g}_{\mu\nu},f^\alpha,f^\alpha_{|\mu},...)(-\tilde{g})^{1/2},
\end{eqnarray}
where $k, K$ are the coupling constants for the scalar, vector field,
respectively; $\ell$ is a free scale length related to $a_0$ (c.f
below, after Eq.(\ref{para})); $\sigma$ is an additional non-dynamical
scalar field; ${\cal F}_{\mu \nu} \equiv U_{\mu,\nu} - U_{\nu,\mu}$;
$\lambda$ is a Lagrange multiplier implementing the constraint
$g^{\alpha\beta}U_{\alpha}U_{\beta}=-1$, which is completely fixed by
variation of the action; the function $F(kG\sigma)$ is chosen to give
the correct non-relativistic MONDian limit, with $G$ related to the
Newtonian gravitational constant, $G_N$, by
$G=G_N/(1+\frac{K+k}{2}-2\phi_c)$, where $\phi_c$ is the present day
cosmological value of the scalar field. Covariant derivatives denoted
by $|_{\mu\dots}$ are taken with respect to $\tilde{g}_{\mu\nu}$ and indices are
raised/lowered with the metric $g_{\mu\nu}$. A new
function $\mu (y)$ is introduced as~\cite{bek}
\bea
-\mu F(\mu ) -\frac{1}{2} \mu^2 \frac{d\,F(\mu)}{d\,\mu} = y \equiv
k\ell^2(g^{\mu\nu}-U^\mu U^\nu)\phi_{,\mu}\phi_{,\nu} \nonumber
\\ kG\sigma^2 = \mu \left(k\ell^2(g^{\mu\nu}-U^\mu
U^\nu)\phi_{,\mu}\phi_{,\nu}\right).
\label{mudef}
\eea
Possible choices for the $\mu$ function will be discussed later.

The isotropic spherically symmetric Einstein metric can be generically
written as $g_{\alpha\beta}{\rm d}x^\alpha {\rm d}x^\beta = -e^\nu
{\rm d}t^2 + e^\zeta({\rm d}r^2+r^2{\rm d}\theta^2+r^2\sin^2\theta
{\rm d}\varphi^2)$; both $\nu$ and $\zeta$ are functions of $r$. The
physical metric $\tilde{g}_{\alpha\beta}$ has the same form, with
$e^{\tilde{\nu}}$ and $e^{\tilde{\zeta}}$, related to the Einstein
metric functions through $\tilde{\nu}=\nu+2\phi$,
$\tilde{\zeta}=\zeta-2\phi$. Isotropy implies $\phi =
\phi(r)$. Assuming an ideal pressureless matter fluid,
$\tilde{T}_{\alpha\beta}=\tilde{\rho}\tilde{u}_\alpha\tilde{u}_\beta$. Motivated
by a homogeneous and isotropic cosmology, the vector field is
considered time-like. The normalisation condition imposed by the
Lagrange multiplier in Eq.~(\ref{teveslagr}) gives
$U^{\alpha}=(e^{-\nu/2},0,0,0)$.

Previous attempts~~\cite{Feix,Feix2,zhao06} at lensing analysis in
TeVeS have remained non-relativistic, considering only the effect of
adding a scalar potential to the standard Newtonian potential.  Hence they
are insensitive to any unique features of TeVeS as a fully
relativistic field theory. The analysis given here is used to solve
(numerically) the TeVeS equations of motion and obtain explicit
expressions for the physical metric quantities $\tilde{\nu}$,
$\tilde{\zeta}$ and thus derive the modified Birkhoff's theorem for
the TeVeS theory. These functions are used to obtain the deflection
angle and find the lensing mass, which is then compared to the stellar
mass content. Assuming a mass density profile $m_{\rm s}(< r)$, within a radial distance $r$, leads
to a system of differential equations to determine $\zeta$ and
$\nu$. A transformation to $\tilde{\nu}$, $\tilde{\zeta}$ then gives
the \textit{physical} metric. The $(tt)$ and $(\theta\theta)$ differential
equations are:
\bea
\label{finalsystem}
&&\zeta''+\frac{(\zeta')^2}{4}+\frac{2\zeta'}{r}+e^{\zeta}\Lambda =
-\frac{kG^2m_{\rm
    s}^2}{4\pi\mu(y)}\frac{e^{-(\nu+\zeta)}}{r^4}\nonumber\\
&&\ \ \ \ \ -\frac{2\pi\mu^2(y)}{l^2k^2}F(\mu)e^\zeta-K
\left[\frac{(\nu')^2}{8}+\frac{\nu''}{2}+\frac{\nu'\zeta'}{4}
+\frac{\nu'}{r}\right]\nonumber\\ &&
\ \ \ \ \ \ \ \ \ \ \ \ \ \ \ \  \ \ \ \ \ \ \ \ \ \ \ \ \ -
8\pi G\tilde{\rho}e^{\zeta-2\phi}~,
\nonumber\\ &&\frac{(\nu'+\zeta')}{2r}+\frac{(\nu')^2}{4}
+\frac{\zeta''+\nu''}{2}+e^{\zeta}\Lambda
= \nonumber \\ && \ \ \ \ \ -\frac{kG^2m_{\rm s}^2}{4\pi
  \mu(y)}\frac{e^{-(\nu+\zeta)}}{r^4}-\frac{2\pi\mu^2(y)}{l^2k^2}
F(\mu)e^\zeta+\frac{K}{8}\nu'^2;
\eea
where the prime denotes derivatives with respect to $r$ and the mass
profile is $m_{\rm s}(<r) =
4\pi\int^r_0\tilde{\rho}\,{\rm exp}\left({\frac{\nu}{2} +\frac{3\zeta}{2}-2\phi}
r^2 \right){\rm d}r$. The deflection angle reads~\cite{bek}:
\bea\label{angle}
\Delta\phi=2\int^\infty_0\frac{1}{r}\left(e^{\tilde{\zeta}-\tilde{\nu}}
\frac{r^2}{b^2}-1\right)^{-1/2}{\rm d}r -\pi~;
\eea
$b^2 = e^{\tilde{\zeta}(r_0)-\tilde{\nu}(r_0)}r_0^2$ is the observable
impact parameter and $r_0$ is the point of closest approach for the
light ray.

The Navarro-Frenk-White (NFW) profile~\cite{NFW} is used, which in
Schwarzschild radial coordinates $\hat{r}$, with
$\hat{r}=e^{\tilde{\zeta}/2}r$, reads
$M(<\hat{r})=M\left[\ln\left(1+\frac{{\cal C}\hat{r}} {r_{\rm
      vir}}\right)-\frac{{\cal C}\hat{r}}{r_{\rm vir}+{\cal
      C}\hat{r}}\right] \left[\ln(1+{\cal C})-\frac{{\cal C}}{1+{\cal
      C}}\right]^{-1}$, where the concentration ${\cal C}$ is ${\cal C}\sim
10$ and $M$ is the total mass of the galaxy contained within the
virial radius, $r_{\rm vir}$; it also specifies the density profile.

Making the approximation $m_{\rm s}(<r)\approx M(<r)$,
shown~\cite{bek} to be correct to leading order of $r$,
Eq.~(\ref{finalsystem}) is numerically solved and through a
transformation to the physical metric, the deflection angle
Eq.~(\ref{angle}) is obtained.

A specification for the function $\mu$ (see, Eq.~\ref{mudef}) is
required.  Previous lensing studies~\cite{Feix2,zhao06} on the
non-relativistic scalar potential approach to TeVeS, adopted a model
for $\mu(y)$ given in Ref.~\cite{bek}.  It was however
noted~\cite{mu1,mu2} that when this choice is converted into its
MONDian equivalent, rotation curve data are not well fitted.  It was
hence proposed to consider~\cite{mu3} the MOND function which best
fits the rotation curve data and then convert it into its TeVeS
$\mu(y)$ analogue. Any intermediate choice for $\mu(y)$ can be
parametrised by $\alpha$.  In what follows the results from testing
this parametrised $\mu(y)$ in the fully relativistic treatment of
lensing in TeVeS are presented in order to see if TeVeS can both fit
rotation curve data and lensing data for a single choice of
$\mu(y)$. For TeVeS $\mu(y)$, $F(\mu)$ and MOND $f(x)$ we use:
\bea
\mu(y)&=&\frac{\sqrt{y/3}}{1-\frac{4\pi\alpha}{k}\sqrt{y/3}} \nonumber\\
F(\mu)&=&\frac{6k^3}{(4\pi\alpha)^3\mu^2}
\left[\ln(\frac{4\pi\alpha\mu}{k}+1)^2+\frac{1}{1+\frac{4\pi\alpha\mu}{k}}
-\frac{4\pi\alpha\mu}{k}\right]\nonumber
\\ f(x)&=&\frac{2x}{1+(2-\alpha)x+\sqrt{(1-\alpha
    x)^2+4x}}~,
\label{para}
\eea where the parameter range is $0<\alpha\leq1$~\cite{mu3}. The
$\alpha = 0$ case for $\mu(y)$ gives the weak and intermediate gravity
limit of $\mu(y)$ and $F(\mu)$ is taken from its explicit
form~\cite{bek}. The $\alpha = 1$ gives the function which better fits
rotation curve data. Since the functions increase monotonically, the
analysis could be confined to the extremes of the parameter space,
i.e. the $\alpha = 0$ and $\alpha = 1$ cases.  The TeVeS parameters
are~\cite{Feix} $k = 0.01,~K = 0.01,~\ell = \sqrt{k\tilde{b}}/(4\pi\Xi
a_0),~\phi_{\rm c} = 0.001,~\Xi=1+ K/2-2\phi_c$.

\begin{figure}[t]
  \centering
\includegraphics[trim = 6mm 6mm 7mm 7mm, clip,scale =.8]{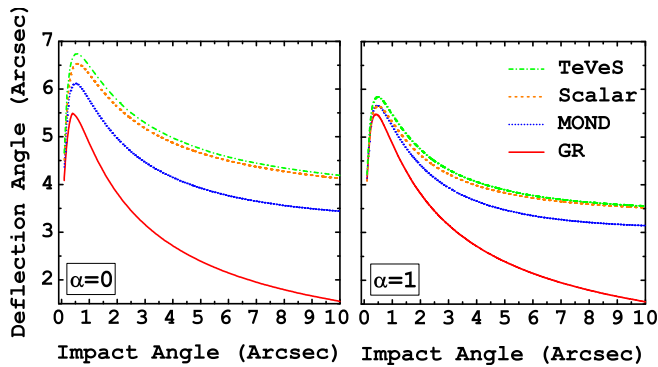}
  \caption{\small{Deflection angle curves for a NFW profile system
      (${\cal C}=10$). The parameters of the system are
      $D_l=1000\ {\rm Mpc}$ (the luminosity distance of the lens),
      $M=10^{11}M_\odot$, $k=0.01, K=0.01, \phi_{\rm c}=0.001$ and
      $r_{vir}=15\ {\rm Kpc}$}. {\sl Left panel:} $\alpha=0$
    case. {\sl Right panel:} $\alpha=1$ case. The deflection angles
    decrease in MOND and TeVeS as $\alpha$ increases.}
  \label{fig:m}
\end{figure}

The cosmological model used is $(\Omega_{\rm
  m},\Omega_\Lambda,\Omega_{\rm k})=(0.3,0.7,0)$, though the effect of
other choices is also considered. The parameter $\tilde{b}$ is fixed
from the limit of the $y(\mu)$ function as $\mu$ becomes $<<1$, 
$y(\mu) \approx \tilde{b}\mu^2$. For our parametrised choice of $\mu(y)$
$\tilde{b}=3$, specifying then $l$. The deflection angle for a model
system is then calculated with this choice of parameters. The results
for the deflection angle in TeVeS are compared to the deflection angle
resulting from following the scalar potential method employed in
Refs.~\cite{Feix,Feix2,zhao06}, as well as GR and MOND. The deflection
angle results are shown in Fig.~\ref{fig:m}.

To calculate a possible dark matter component, a sample of double
lensing systems from the Castles database is analysed and the mass of
the lensing galaxy in GR, MOND, and TeVeS is calculated. By comparing
the mass from lensing to the stellar mass content calculated from a
comparison of photometry and stellar population synthesis using a
Chabrier Initial Mass Function (IMF), as in Ref.~\cite{fsw05}, the
mass deficit which belonged to the ``dark'' sector is found. Note that
while the authors of Ref.~\cite{sand08} claim that the stellar mass estimates are
sensitive to the IMF used, it has been argued in Ref.~\cite{FSans} that the main
competitor of the Chabrier IMF, the Salpeter IMF, appears to fit the data worse. 
Other realistic IMF choices differ from the Chabrier by an
insignificant factor~\cite{FSans}. These arguments support the
validity of the method employed here. Finally, inverse ray tracing is
used to calculate the mass~\cite{fsy}. The mass estimates we obtain this way 
are shown in Table 1.

\begin{table}[ht]

\begin{center}
\begin{tabular}{l||r|rr|rr||c}
\hline & \multicolumn{1}{c}{GR} & \multicolumn{2}{c}{MOND} &
\multicolumn{2}{c}{TeVeS} & {M$_{\rm STAR}$}\\ \textbf{Lens} & &
$\alpha = 0$ & $\alpha=1$ & $\alpha=0$ & $\alpha=1$ &
\multicolumn{1}{c}{Ref.~\cite{fsw05}}\\ \hline
\scriptsize{HS0818+1227} & $34.9$ & $23.9$ & $27.6$ & $19.1$ & $24.4$
& $16.2^{21.2}_{12.6}$\\ \scriptsize{FBQ0951+2635} & $3.1$ & $2.3$ &
$2.6$ & $1.9$ & $2.3$ & $1.1^{2.1}_{0.5}$\\ \scriptsize{BRI0952-0115}
& $3.6$ & $2.4$ & $2.8$ & $1.9$ & $2.4$ &
$3.5^{4.0}_{2.7}$\\ \scriptsize{HE1104-1805} & $86.1$ & $59.0$ &
$68.5$ & $47.1$ & $60.4$ &
$22.8^{51.2}_{12.7}$\\ \scriptsize{LBQS1009-0252} & $15.2$ & $10.7$ &
$12.2$ & $8.7$ & $10.9$ & $5.5^{7.9}_{4.2}$\\ \scriptsize{HE2149-2745}
& $11.2$ & $7.8$ & $9.0$ & $6.3$ & $8.0$ & $4.6^{6.7}_{3.6}$\\ \hline
\end{tabular}
\end{center}
\label{table:mass}
\caption{Mass estimates (in $10^{10}M_\odot$ units) for $\Lambda$CDM
  cosmology: $(\Omega_{\rm m},\Omega_\Lambda,\Omega_{\rm
    k})=(0.3,0.7,0)$, $k=K=0.01$, and a NFW halo profile with ${\cal
    C}=10$. Two different cases ($\alpha=0,1$) for $\mu(y)$
  parametrisation are considered.  }
\end{table}

\begin{figure}[!ht]
  \includegraphics[trim = 7mm 6mm 8mm 8mm, clip,scale=0.95]{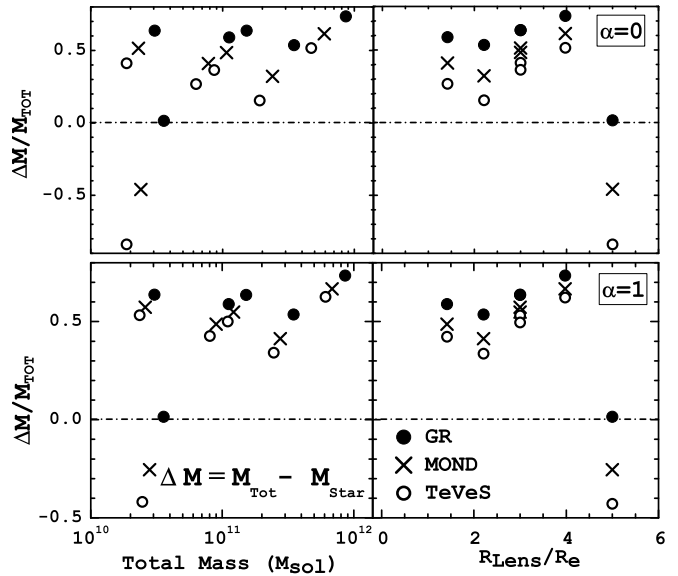}
  \caption{\small{A comparison of the need for dark matter for the
      lenses in GR, MOND and TeVeS, for $\alpha = 0$ (top
      panels) and $\alpha = 1$ (bottom panels). The right hand panels
      give the dark matter requirements as a function of the ratio
      $R_{lens}/R_e$. Any dark matter needed in TeVeS is especially
      significant for those masses which probe further out, where the
      modification to gravity is larger.}}
  \label{fig:diff}
\end{figure}

Figure \ref{fig:diff} compares the mass estimates between GR, MOND and
TeVeS for the two cases of $\mu(y)$, with the mass difference being
given as a function of the mass calculated using GR (left panels) and
$R_{\rm lens}/R_{\rm e}$ (right panels), where $R_{\rm lens}$ is the
distance out to which our mass estimates and the stellar mass
estimates are calculated and $R_{\rm e}$ is the half light radius,
both given in Ref.~\cite{fsw05}. The top two panels are for $\alpha =
0$, the bottom two are for $\alpha = 1$. Comparing the lensing masses
against the stellar masses (excluding one outlier), it is found that
even in TeVeS on average the dark matter content is 48.5$\%$ when
$\alpha =0$, and 34.3$\%$ when $\alpha=1$. The outlier lensing system
BRI0952-0115 shows a mass overshoot of $\approx80\%$ in TeVeS,
i.e. the lensing mass in TeVeS is less than the stellar mass
content. It is possible that this system is affected by some unknown
lens environment effects such as an unseen cluster mass contribution
as has been suggested of other lenses~\cite{Feix2}, though there is no
data to conclude this at present. Overall the analysis shows that
TeVeS finds it hard to explain the lensing observed in these systems
using only the stellar content of the galaxies, a conclusion which
stands in contrast to that given in Refs.~\cite{Feix2,zhao06} despite
the two methods predicting similar deflection angles (see,
Fig.\ref{fig:m}). This is a problem for TeVeS, which will be made more
explicit later on in this letter, when the analysis of varying the
free parameters will be performed.

The results show that the case $\alpha = 1$, which is a choice
specifically adopted to fit the rotation curve data, requires an even
larger amount of dark matter.  Thus, attempts to fit TeVeS to a no
dark matter scenario using the freedom in the parameters of the class
of $\mu$ functions used here and in the literature so far, would imply
that the theory fits poorly the rotation curves data, which was the
original motivation for modifying the gravitational behaviour.

The effect of different cosmological parameters has also been examined
to see how they alter the results for the $\alpha =0$ case,
corresponding to the minimum amount of required dark matter.  It is
found that the lensing results are insensitive to the precise
cosmological parameters, which is not surprising, since the
observational constraints mostly impose limits on the luminosity and
angular distance scales~\cite{fsy}.  For completeness we state the
results, all of which point towards fluctuations in the amount of dark
matter well within the error limits. In particular, for the
case~\cite{zhao06} $(\Omega_{\rm m},\Omega_\Lambda,\Omega_{\rm
  k})=(0.03,0.46,0.51)$, averaging over the six galaxies, we find a
5.7\% increase in the amount of dark matter required. For the case
$(\Omega_{\rm m},\Omega_\Lambda,\Omega_{\rm k})=(0.23,0.78,0)$,
considered in Ref.~\cite{sko06}, to fit the Cosmic Microwave
Background data with TeVeS, one finds a corresponding average decrease
by 2.2 \%, while for $\phi_c=0.01$, one finds an average decrease by
1.1\%~.

Finally, the $k,~K$ parameters are varied independently to check on
the robustness of our claims. In particular, we examine the lensing
system HE1104-1805, which within the TeVeS approach (with $\alpha=0$)
requires the largest amount of dark matter. We find
that the variation of the parameter $k$ has considerably smaller
effects than that of $K$. This supports the dominant r\^ole played by
the vector field in TeVeS. A similar result has also been pointed out
in Ref.~\cite{liguori}, but from a different perspective.  Within the
allowed parameter space $1\leq K\leq 10^{-5}$, $1\leq k\leq 10^{-5}$,
implied by rotation curve data and solar system tests of gravity, the
amount of dark matter required in the system HE1104-1805 is negligible
only for values of $K$ higher than 0.1, which however is excluded by
gravitational measurements at solar-system scales.

\begin{figure}[!ht]
\centering
\includegraphics[trim = 7mm 7mm 6mm 9mm, clip,scale =0.85]{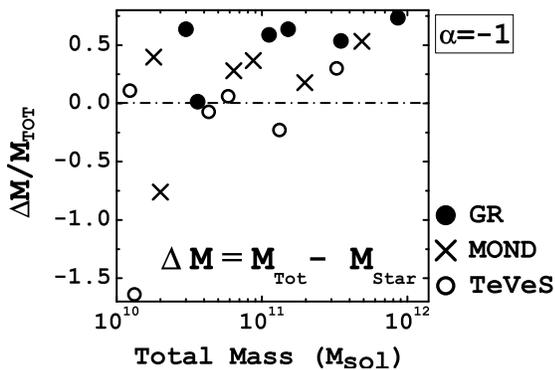}
\caption{\small{The dark matter content of the lenses for $\alpha =
    -1$ in the parametrisation of $f(x)$ in MOND and $\mu(y)$ in
    TeVeS.}}
\label{fig:par}
\end{figure}

We also considered other values of $\alpha$, outside the region
$\alpha\in [0,1]$, considered in the literature so far.  We found that
the $\alpha = -1$ case appears not to require substantial amount of
dark matter to explain the gravitational lensing, and hence it could
provide an example of an altrernative to the dark matter
scenario. However, such a model is in conflict with data from rotation
curves of galaxies.

In conclusion, the above results show a soft dependence on the free
parameters of TeVeS and the cosmological model adopted, but a rather
strong one on the form of the $\mu(y)$ function. Our analysis in this
letter points towards the fact that TeVeS, at least within the class
of models considered so far in the literature, cannot survive both
gravitational lensing and rotation curve tests. However, we cannot yet
exclude completely the possibility, admittedly remote, that a class of
$\mu$ functions, or more complicated equivalents thereof, can be found
such that alternative to dark matter scenarios are at play.

\acknowledgments It is a pleasure to thank Ignacio Ferreras for
discussions. The work of N.\,E.\,M. and M.\,S. is partially supported
by the European Union through the Marie Curie Research and Training
Network UniverseNet (MRTN-CN-2006-035863), while that of M.F.Y. is
supported by an E.P.S.R.C. (UK) studentship.

\end{document}